\documentclass[twocolumns]{aa}
\usepackage{graphicx}
\usepackage{txfonts}
\usepackage{natbib}
\begin{document}
   \title{The HELLAS2XMM survey}

   \subtitle{IX. Spectroscopic identification of super-EROs hosting
         AGNs 
	  \thanks{Based on data obtained at the VLT through
	 the ESO program 73.A-0598(A).}
	}

   \author{R.~Maiolino\inst{1}
          \and
          M.~Mignoli\inst{2}
	  \and
	  L.~Pozzetti\inst{2}
	  \and
	   P.~Severgnini\inst{3}
	  \and
	  M.~Brusa\inst{4}
	  \and
	  C.~Vignali\inst{5}
	  \and
	  S.~Puccetti\inst{6} 
	  \and
	  P.~Ciliegi\inst{2}
	  \and
	  F.~Cocchia\inst{6}
	  \and
	  A.~Comastri\inst{2}
	  \and
	  F.~Fiore\inst{6}
	  \and
	  F.~La~Franca\inst{7}
	  \and
	  G.~Matt\inst{7}
	  \and
	  S.~Molendi\inst{8}
	  \and
	  G.C.~Perola\inst{7}
          }

   \offprints{R. Maiolino}

   \institute{INAF - Osservatorio Astrofisico di Arcetri,
      L.go E. Fermi 5, 50125 Firenze, Italy\\
              \email{maiolino@arcetri.astro.it}
         \and
             INAF - Osservatorio Astronomico di Bologna, via Ranzani 1,
	     	40127 Bologna, Italy\\
             \email{marco.mignoli,lucia.pozzetti,andrea.comastri@bo.astro.it}
         \and
	 INAF - Osservatorio Astronomico di Brera 28, 20121 Milano, Italy\\
             \email{paola@brera.mi.astro.it}
	  \and
             Max-Planck-Institut f\"{u}r Extraterrestrische Physik, Garching,
	       Germany\\
             \email{marcella@mpe.mpg.de}
         \and
             Dipartimento di Astronomia, Universit\'a di Bologna, via Ranzani 1,
	     	40127 Bologna, Italy\\
             \email{cristian.vignali@bo.astro.it}
         \and
             INAF - Osservatorio Astronomico di Roma, via Frascati 33, 00040
	      Monteporzio, Italy\\
             \email{cocchia,fiore,puccetti@mporzio.astro.it}
         \and
              Dipartimento di Fisica Universit\'a di Roma Tre, via della
	      Vasca Navale 84, 00146 Roma, Italy\\
             \email{lafranca,matt,perola@fis.uniroma3.it}
         \and
              IASF-CNR, Istituto di Fisica Cosmica, via Bassini 15, 20133
	      Milano, Italy\\
             \email{silvano@mi.iasf.cnr.it}
             }

   \date{Received ...; accepted ...}

   \abstract{
   We present VLT near-IR
   spectroscopic observations of three X-ray
   sources characterized by extremely high X-ray--to--optical
ratios (X/O$>$40), extremely red colors (6.3$<$R--K$<$7.4, i.e. EROs)
and bright infrared magnitudes (17.6$<$K$<$18.3). These objects
are very faint in the optical, making their spectroscopic identification
extremely challenging. Instead, our near-IR spectroscopic observations
have been successful in identifying the redshift of two of
them (z=2.08 and z=1.35), and tentatively even of the third one
(z=2.13).
When
combined with the X-ray properties, our results clearly indicate that all these
objects host obscured QSOs
($\rm 4\times10^{44}<L_{2-10keV}<1.5\times 10^{45}~erg~s^{-1}$,
$\rm 2\times10^{22}<N_H <4\times 10^{23}cm^{-2}$) at high redshift.
The only object with unresolved morphology in the K band shows
broad H$\alpha$ emission,
but not broad H$\beta$, implying a type 1.9 AGN classification.
The other two objects are resolved and dominated by the host galaxy
light in the K band, and appear relatively quiescent:
one of them has a LINER-like emission line spectrum
and the other presents only a single, weak emission line which we tentatively
identify with H$\alpha$.
The galaxy luminosities for the latter two objects are an order of magnitude
brighter than typical local L$_K^*$ galaxies and the derived
stellar masses are well in excess of $\rm 10^{11}~M_{\sun}$. For these
objects we estimate black hole masses higher than $\rm 10^9~M_{\odot}$
and we infer that they are radiating at Eddington ratios
$\rm L/L_{Edd}\le 0.1$. We discuss the implications of these
findings for the coevolution of galaxies and black hole growth.
Our results provide further support that X-ray sources
with high X/O ratios and very red colors tend to host obscured QSO in
very massive galaxies at high redshift.

   }

   \maketitle
%

\section{Introduction}

During the past few years hard X-ray surveys have clearly revealed the
important role played by obscured AGNs for the cosmic
X-ray background and for the accretion history of supermassive
black holes \citep[][]{brandt01,giacconi02,hasinger01,fiore03,ueda03,
marconi04}.
Obscured AGNs are found to be 3--4 times more numerous
than unobscured AGNs. This figure seems to decrease at higher,
QSO-like luminosities \citep[][]{fiore03,ueda03,lafranca05}, although
this trend has been recently questioned by \cite{treister05}.
One of the main issues affecting these studies (and in particular
the obscured--to--unobscured QSO ratio) is that
a significant fraction of high-z, luminous obscured
AGNs may have escaped optical spectroscopic identification due to the
weakness of their optical counterparts.
In most of these cases multiband photometry
is the only viable resource to constrain their redshift.

Within this context a new interesting class of objects, emerging
from the X-ray surveys, are sources with very high X-ray--to--optical ratio
(hereafter X/O)\footnote{$\rm X/O = F_X/F_{opt}=  F_X /10^{-(0.4~R+5.4)}$,
where $\rm F_X$ is the observed X-ray flux in the band 2--10~keV
(in units of $\rm erg~s^{-1}~cm^{-2}$) and
R is the R-band magnitude.},
and in particular those with X/O$>$10 (to be compared with the value of
X/O$\sim$1 of unobscured, type 1 AGNs). In deep surveys
($\rm F_{2-10keV}<10^{-14}erg~cm^{-2}s^{-1}$) very few sources
with X/O$>$10 have been identified spectroscopically \citep[][]{barger03,
mainieri05}, since the combination of low X-ray fluxes and high X/O
results into optical magnitudes R$>$25, extremely difficult to observe
spectroscopically even with the largest telescopes currently available.
Instead, large area shallower surveys, such as the
HELLAS2XMM survey ($\rm F_{2-10keV}>10^{-14}erg~cm^{-2}s^{-1}$
over 1 sq. degree),
delivered samples of high-X/O sources with
brighter optical counterparts, therefore suitable for spectroscopic
follow-up. In particular, \cite{fiore03} have spectroscopically identified
a sizable sample (13 sources) of HELLAS2XMM sources with X/O$>$10 and
R$\sim$24, many of which are type 2 QSOs at high redshift.

However, a subsample of these HELLAS2XMM sources with
extreme X/O ratios (30$<$X/O$<$150)
are difficult to identify spectroscopically even within the HELLAS2XMM
sample, due to the extremely faint optical magnitudes of their
counterparts. \cite{mignoli04}
obtained near-IR, K-band images of 10 sources characterized by
X/O$>$30 and R$>$24.5. The most surprising finding were the bright
near-IR magnitudes of these objects,
in the range 17.6$<$K$_s$$<$19.1, resulting into
colors R--K$>$5, placing them into the class of Extremely Red Objects
(EROs). Thanks to excellent seeing conditions, most
of the sources were resolved, showing elliptical-like profiles in most
cases (but two point-like sources and one disky-profile are also present).
Colors and sizes suggest redshifts larger than about 1. When combined
with the X-ray fluxes and slopes, their results suggest that these sources
host type 2 QSOs, whose light is totally absorbed in the optical and
often even in the near-IR.
Such objects with extreme X/O ratios and extremely red colors probably represent
10\% of the sources with $\rm L_X > 10^{44} erg~s^{-1}$ in the HELLAS2XMM
sample \citep[][]{mignoli04}.
Obviously, a spectroscopic investigation and redshift confirmation
of these sources is highly desirable, and would help us to tackle the
following issues: 1) confirm the QSO2 nature of the ``superEROs'' in the
\cite{mignoli04} sample; 2) determine their contribution to the census
of obscured AGNs at high redshift and compare it
with the models of the X-ray background
\citep[][]{perola04,fiore03,mainieri05,lafranca05}; 3) investigate further
the trends of the AGN2/AGN1 ratio as a function of luminosity and redshift
\citep[][]{fiore03,ueda03,lafranca05};
4) confirm that these X-rays sources are hosted in very massive galaxies
at high redshift, and investigate the implications for the QSO-spheroids
coevolution.

The relatively bright magnitudes characterizing the near-IR counterparts
of the X-ray sources in the \cite{mignoli04} sample
suggest that near-IR spectroscopy may be a more effective
tool to investigate these extremely red sources than optical spectroscopy.
Moreover, the rest-frame optical AGN lines could emerge in the near-IR, since
dust extinction is reduced with respect to the rest-frame UV lines.
We have performed a pilot program of near-IR spectroscopy of four HELLAS2XMM
sources with R--K$>$6, although we could obtain a full near-IR
spectrum only for three of them. As discussed in this paper, the results are
extremely encouraging, with the determination of secure
spectroscopic redshifts for two sources and a tentative redshift for the third
one.

In Sect.~2 we describe the sample selection, along with the observations
and the data reduction process. In Sect.~3 we present the resulting spectra
and discuss each source individually. In Sect.~4.1 we discuss the implications
of our results for the connection between QSO2, EROs and high X/O sources.
In Sect.~4.2 we infer the stellar masses of the host galaxies, the black
hole masses and the accretion rates in terms of Eddington luminosity.
Finally, in Sect.~5 we draw the main conclusions.

In this paper we assume the concordance $\Lambda$-cosmology with
$\rm H_0=71~km~s^{-1}Mpc^{-1}$, $\rm \Omega _{\Lambda}=0.73$ and
$\rm \Omega _{m}=0.27$ \citep[][]{spergel03}.

\begin{table}[bh!]
\label{tab1}      
\caption{Summary of the targets properties}             
\centering          
\begin{tabular}{l c c c l }     
\hline\hline       
Source  &  X/O & K$_s$ & R--K$_s$ & Morph.(K)\\ 
\hline                    
Abell2690\#029 &  78 & 17.68 & 7.4$\pm$1.0 & Point. \\ 
BPM16274\#069  &  52 & 17.83 & 6.6$\pm$0.8 & Ell./Disc. \\ 
Abell2690\#075 &  45 & 18.32 & 6.3$\pm$0.7 & Ellip. \\ 
PKS0537-28\#111& 154 & 17.66 & 6.8$\pm$0.7 & Ellip. \\
\hline                  
\end{tabular}
\end{table}


   \begin{figure}[h!]
   \centering
   \includegraphics[width=9cm]{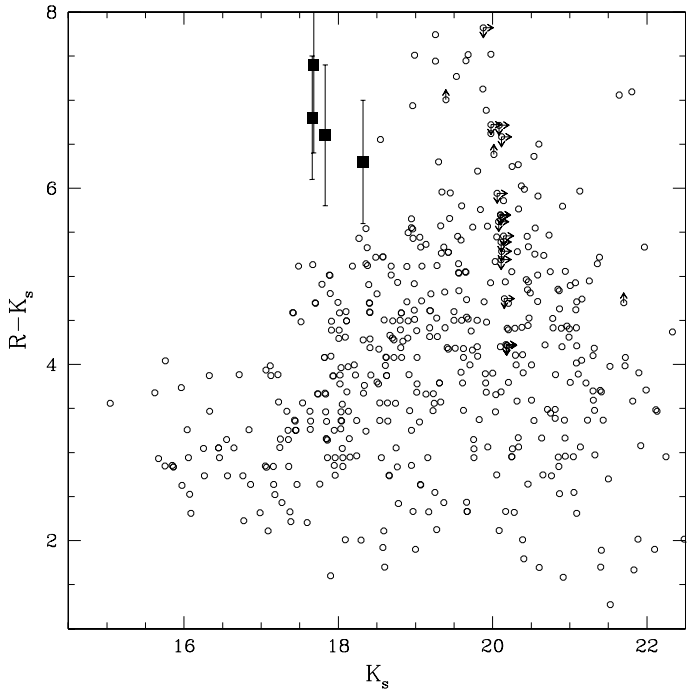}
      \caption{Comparison of the K-band fluxes and R-K colors of the sources
      in our sample (large solid symbols) with the X-ray
      sources in the deep, pencil
      beam survey in the Chandra Deep Field North \citep[small empty
      symbols][]{barger03}. Note the extreme properties of our
      objects in terms of brightness and
      colors which sample a region totally unexplored by the deep surveys.
              }
         \label{fig_krk}
   \end{figure}

\section{Sample selection, observations and data reduction}

Out of the 11 X-ray HELLAS2XMM high X/O sources
in the \cite{mignoli04} sample (10 of which have EROs colors),
we selected four sources with
6.3$<$R--K$<$7.4, K$<$18.3, X/O$>$40 and different morphological types.
Table 1 summarizes the properties of the sources in our pilot program.
Fig.\ref{fig_krk} shows the location of our four sources in a K vs. R--K
diagram (big squared
dots). The figure
highlights the extreme properties of our ``super-EROs'', in terms of
colors and brightness, relative to the population of X-ray sources found
in deep pencil surveys, such as the CDFN \citep[][]{barger03},
whose distribution is indicated
with small symbols\footnote{\cite{barger03} list the photometry
in the HK' filter. We converted this photometry into K band by using
the relation $\rm HK'-K = 0.13+0.05(I-K)$, provided in the same paper.}.
Other shallow surveys have found X-ray sources with near-IR counterparts
with similar extreme properies \citep[][]{gandhi04,severgnini05,brusa05}.
However, within this class of X-ray super-EROs the ones presented in this
paper are among the few ones with rest-frame optical spectra which allow,
beside the redshift determination, to investigate their optical
properties.

Observations
were obtained with ISAAC at ESO-VLT during period 73 (April 2005 --
September 2005) in service mode. We used the
low resolution mode with 1$''$ slit (R$\sim$500) in the
J (1.16--1.34 $\mu$m), H (1.48--1.76 $\mu$m) and
K (1.97--2.5 $\mu$m) bands. PKS0537\#111 is the only object
which was observed only in the H band.
For each band we used eight integrations
of 300 seconds each, and moving the target along the slit
(following an A-B-B-A pattern),
for a total of 40 minutes of integration per band. In some cases the
observation was repeated due to the poor seeing in the first observation
($>1.2''$). 

Data reduction followed the standard threads. The background was
removed by subtracting contiguous A-B frames. Flat fielding was
performed by using the master flat provided by the ESO pipeline.
Spectroscopic calibration was obtained through the spectrum of an
arc (Ar-Xe) lamp. Small wavelength shifts
due to the grating positioning uncertainty were corrected by means
of the OH sky lines. The individual frames were aligned and averaged,
with a sigma-clipping threshold to remove bad pixels and cosmic
rays. Telluric absorption features and the relative response of the
instrument were corrected by dividing the spectrum by a stellar standard
of known spectral type\footnote{The intrinsic shape of the standard
stars was corrected by using 
the spectral libraries of \cite{pickles98} or through the
method discussed in \cite{maiolino96} in the case of late type main sequence
stars.}. The absolute flux
calibration (and therefore also inter-band calibration) is
problematic. Indeed, we do not have photometry for the
targets in the J and H bands
which could be used to carefully calibrate the spectra. So we attempted
to calibrate the spectra by using the standard star taken with the slit
of 2$''$ and trying to estimate the slit losses through the seeing observed
in the acquisition images. We checked that in the K band (where we have
photometry) the accuracy of the flux calibration is about 25\%, but we
do not have any control for the J and H bands.
Moreover, in the case of BPM16724\#069 no standard
star was taken, so we had to rely on standard stars taken 10--20 nights
earlier, implying larger uncertainties both for the absolute flux calibration
and even for the intra-band slope and relative calibration.

   \begin{figure}[h!]
   \centering
   \includegraphics[width=9cm]{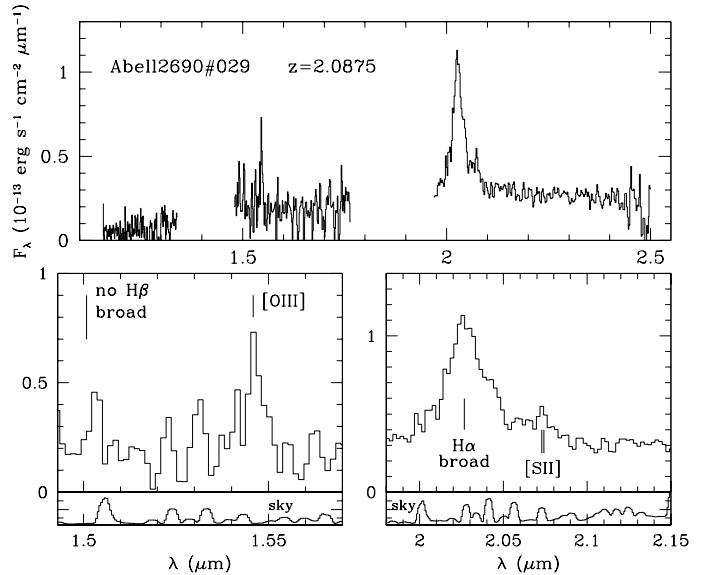}
      \caption{ISAAC-VLT near infrared spectrum of Abell2690\#029,
      a type 1.9, red QSO at z=2.087. The spectrum is smoothed
      with a 6 pixel boxcar. The bottom panels show a zoom of
      the spectrum around [OIII] and H$\alpha$. The lower
      insets show the sky spectrum (arbitrary flux units).
              }
         \label{fig_ab29}
   \end{figure}

   \begin{figure}[h!]
   \centering
   \includegraphics[width=9cm]{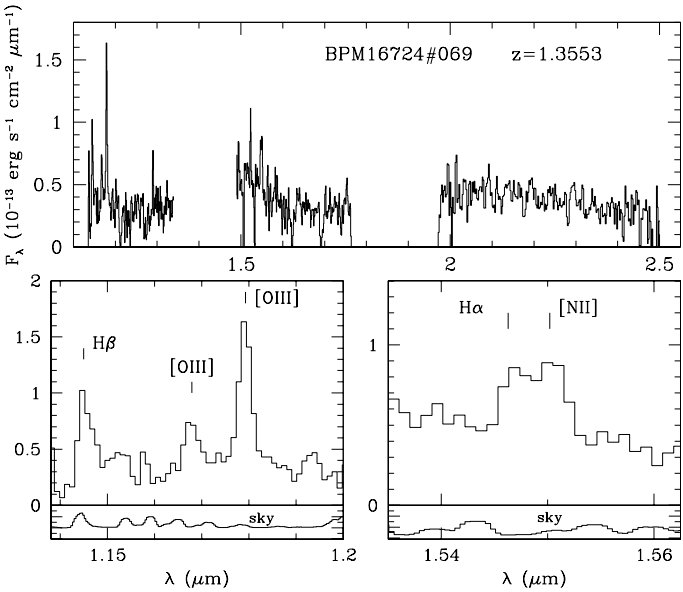}
      \caption{As Fig.\ref{fig_ab29} for BPM16724\#069,
      a LINER-like AGN at z=1.355.
              }
         \label{fig_bp69}
   \end{figure}

   \begin{figure}[h!]
   \centering
   \includegraphics[width=9cm]{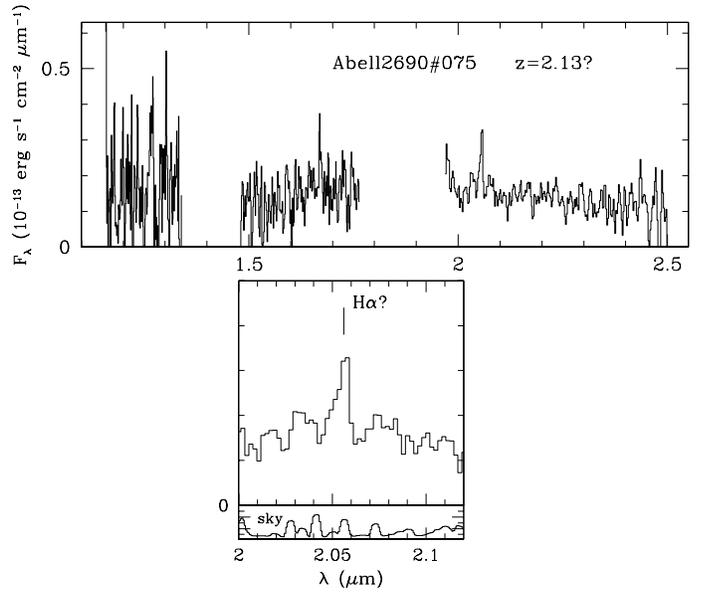}
      \caption{As Fig.\ref{fig_ab29} for Abell2690\#075.
      The bottom panel shows the zoom around a line tentatively
      identified as H$\alpha$ at z=2.13.
              }
         \label{fig_ab75}
   \end{figure}


\section{Results}

The single H-band spectrum of PKS0537\#111 is featureless. The
limited spectral coverage did not allow us to provide useful constraints
on the redshift or on the nature of the source, other than supporting
the stellar nature of the near-IR light inferred from the K band image
by \cite{mignoli04}. This source is no further discussed.

Figs.\ref{fig_ab29}--\ref{fig_ab75} show the resulting spectra of the three sources
observed in all the three bands, smoothed
with a boxcar of 6 pixels (which corresponds to the projected
slit width) to
improve the S/N without significantly affecting the spectral resolution
(the final effective resolution after this smoothing is 420).
Pixels were rebinned to half of the spectral resolution element.
The main observational results are reported in Tab.2.
We discuss each individual object in the following.

\begin{table*}
\label{tab2}      
\centering          
\caption{Main results from the observations}             
\begin{tabular}{l c c c c c c c}     
\hline\hline       
  &  & \multicolumn{4}{c}{Emission lines} &      & \\ 
 \cline{3-6}
Source & z &  Id. & $\lambda _{obs}$ & Flux  & EW(rest)  & $\rm N_H$ & L$_{\rm 2-10keV}$\\
        &   &      &  ($\mu$m)        & 10$^{-16}$ erg cm$^{-2}$ s$^{-1}$ & \AA &
	(10$^{22}$cm$^{-2}$)	& (10$^{44}$erg s$^{-1}$) \\
\hline                    
Abell2690\#029 & 2.0875 & H$\alpha$(broad) & 2.0266$\pm$0.0015 & 20$\pm$4 & 260$\pm$50 & 2.2$^{+2.7}_{-1.7}$ & 9.7 \\
               &        & [OIII]5007       & 1.5457$\pm$0.0008 & 1.2$\pm$0.2 & 37$\pm$7 &     &       \\
               &        & H$\beta$(broad)  & --     & $<$1 & -- &  &        \\
BPM16274\#069  & 1.3553 & [OIII]5007       & 1.1790$\pm$0.0002 & 4.6$\pm$0.5$^a$& 43$\pm$5 & 2.6$^{+1.5}_{-1.0}$ & 3.9 \\ 
               &        & [OIII]4960       & 1.1680$\pm$0.0005 & 1.7$\pm$0.5$^a$& 13$\pm$4 &     &     \\ 
               &        & H$\beta$(narrow) & 1.1446$\pm$0.0005 & 2.1$\pm$0.6$^a$& 21$\pm$6     &     \\ 
               &        & H$\alpha$(narrow)& 1.5463$\pm$0.0005 & 2.3$\pm$0.7$^a$& 10$\pm$3    &     \\ 
               &        & [NII]6583        & 1.5502$\pm$0.0005 & 2.8$\pm$0.7$^a$& 12$\pm$3     &     \\ 
Abell2690\#075 & 2.13$^b$ & H$\alpha$(?)$^b$& 2.0574$\pm$0.0020$^b$ & 0.81$\pm$0.15$^b$ & 30$\pm$6 & 36$^{+47}_{-22}$$^b$ & 15.0$^b$\\ 
\hline                  
\end{tabular}
\begin{list}{}{}
\item[$^a$] Highly uncertain flux calibration.
\item[$^b$] Redshift and other derived quantities
 based on the tentative identification of the feature at 2.0574~$\mu$m with H$\alpha$.
\end{list}
\end{table*}

\subsection{Abell2690\#029}

The spectrum shows a red continuum with a prominent broad H$\alpha$
(FWHM$\sim$4200~km/s)
in the K band, whose identification is supported by the detection
of [OIII]5007\AA \ in the H band. The [SII] doublet at 6717+6731\AA \ is
also detected (any [NII]6584\AA \ 
line is blended with the broad H$\alpha$). These
emission lines imply a redshift of 2.087, close to the
redshift of 2.4 inferred by \cite{fiore03}, based on the X/O--L$_X$
correlation for obscured systems, and by \cite{mignoli04},
based on the colors.
Broad H$\beta$ in the H band is not detected, yielding to a ``type 1.9''
classification for the AGN
\citep[e.g.][]{osterbrock93}. Note that the narrow feature
      near to the expected location of H$\beta$ is probably a bad
      subtraction of the very strong sky line at this wavelength, although
      some contribution from a narrow component of H$\beta$ cannot be excluded.

From the lower limit on the H$\alpha$/H$\beta$ broad lines ratio we can
set a lower limit to the extinction in the range $\rm A_V > 2-5$~mag,
depending on the assumed intrinsic value of the Balmer ratio
\citep[see discussion in][]{maiolino01a} and on the assumed extinction
curve \citep[][]{gaskell04,hopkins04,maiolino01a}.
On the other hand, the detection of a broad H$\alpha$ requires the visual
extinction not to be much higher than $\sim$5~mag, otherwise the corrected
H$\alpha$ flux would imply a ratio L(H$\alpha$)/L$_X \gg 1$, i.e.
much larger than observed even in the most extreme AGNs
\citep[][]{koratkar95}.

At a redshift of 2.087 the fit of the X-ray spectrum gives an
absorbing column density of $\rm 2.2\times 10^{22}~cm^{-2}$
\citep[][]{perola04}
\footnote{The extinction expected from
the X-ray absorbing column density assuming a Galactic gas-to-dust ratio
and extinction curve is higher (A$_V$(exp)$\sim$10~mag) than inferred
from the near-IR spectrum, but marginally consistent given
the errors on $\rm N_H$. Moreover, a mismatch between optical and
X-ray absorption is common to most AGNs \citep[][]{maiolino01b}.}.
The implied 2--10~keV luminosity corrected
for absorption is $\rm \sim 10^{45}~erg~s^{-1}$,
placing this object in the class of obscured QSOs.

\subsection{BPM16274\#069}

A redshift of 1.355 is secured by the detection of [OIII]4960+5007\AA \ and
(narrow) H$\beta$ in the J band
and a noisier detection of (narrow) H$\alpha$ and
[NII]6584 in the H band. Such a redshift is close to the values
inferred both by \cite{fiore03} and \cite{mignoli04}, 1.6 and $>$1.4
respectively, based on the X-ray, optical and near-IR photometric
properties, as discussed above.

The line ratios ($\rm H\beta /[OIII]=1.03$, $\rm [NII]/H\alpha=1.0$)
are in the LINERs range, which is
not unusual even in several obscured, intrinsically powerful 
nearby AGNs \citep[][]{maiolino03}. The ratio of
H$\alpha$/H$\beta$$\sim$1 is unusually low, and probably reflects the large
uncertainties affecting the inter-calibrations of the J and H bands of this
object (Sect.2).

Also for this object the spectroscopic redshift
implies substantial gaseous column density ($\rm \sim 2.2\times 10^{22}~cm^{-2}$)
along the line of sight and
a high intrinsic luminosity (Tab.2), yielding to a type 2 QSO classification.

\subsection{Abell2690\#075}

There is only one tentative line detection at $\sim$2.05 $\mu$m.
If identified
with H$\alpha$ the inferred redshift would be 2.13. This is very close
to the redshift of 1.9 inferred by \cite{fiore03}, based on the
X/O--L$_X$ correlation, and consistent
with the lower limit of 1.30 inferred by \cite{mignoli04}.


Alternatively, the line at 2.05$\mu$m could be [OIII]5007\AA ,
implying a redshift of 3.11. This could be corroborated by a very
marginal detection of a feature at 1.59$\mu$m, which could match
the wavelength expected for [NeIII]3869\AA \ and which often accompains [OIII].
Any other possible identification of the 2.05$\mu$m feature
with other emission lines weaker than H$\alpha$ and [OIII] is unlikely.

Higher signal-to-noise spectra are required to better confirm the 
line(s) detection and the redshift. In this paper we will assume the working
hypothesis that the redshift is 2.13, i.e. assuming that the tentative
line detection is H$\alpha$. Should the line turn out to be [OIII]5007\AA \
from future, higher quality spectra, then our conclusions would be further
reinforced: a higher redshift would imply even higher X-ray luminosities,
higher column densities and higher galaxy light and masses than estimated
in the following of this paper.

The tentative spectroscopic redshift of 2.13
implies substantial absorbing $\rm N_H$ along the line of sight and
a high intrinsic luminosity (Tab.~2), implying a type 2 QSO classification
also in this case. We note that the X-ray spectrum of this source is
so flat that it could also be consistent with a reflection-dominated
Compton thick QSO. In the latter case the intrinsic X-ray luminosity
would exceed $\rm 10^{46}~erg~s^{-1}$. Such an extreme QSO, in terms
of obscuration, would be similar to the one found by \cite{norman02}
in the CDFS, although intrinsically much more luminous.


\section{Discussion}

\subsection{QSO2 in super-EROs}

As outlined in the previous section and in Tab.2, the redshift
estimated through our spectra imply QSO-like luminosities
($\rm L_{2-10keV}>3\times 10^{44}erg~s^{-1}$) and significant gas absorption
along the line of sight ($\rm N_H>2\times 10^{22}cm^{-2}$), i.e.
all three objects host type 2, obscured QSOs. This class of sources
has been long sought in the past. Recent campaigns of optical
spectroscopic
identification of X-ray sources \citep[][]{barger03,fiore03,szokoly04}
are providing more cases of QSO2, but they are still under numerous
with respect to unobscured, type 1 QSOs.

   \begin{figure}[h!]
   \centering
   \includegraphics[width=9cm]{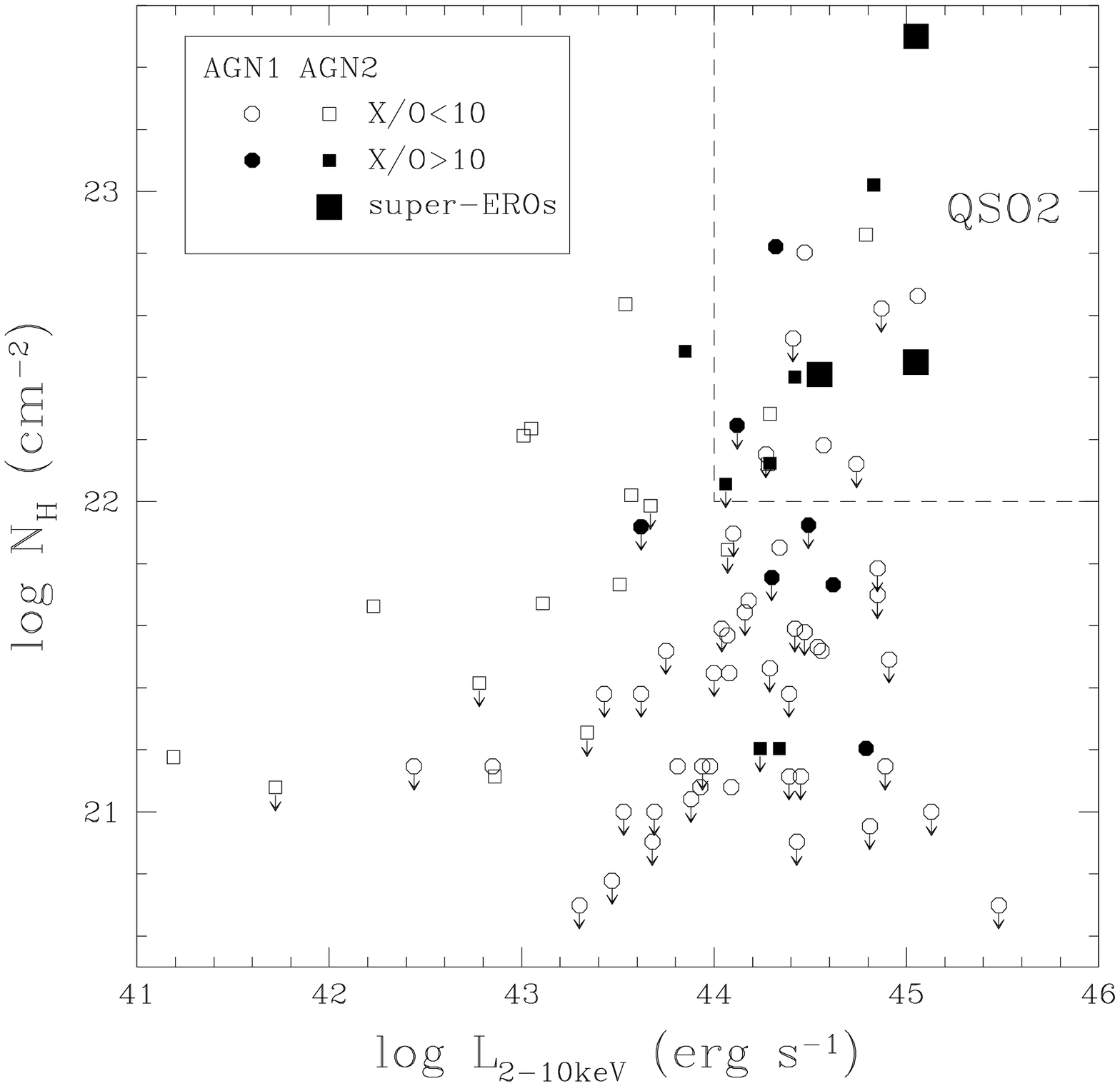}
      \caption{$\rm N_H$ versus (absorption corrected)
      $\rm L_{2-10keV}$ diagram  where the sources
      identified through near-IR spectroscopy in this paper (large symbols)
      are compared with the other HELLAS2XMM
      sources identified through optical spectroscopy
      by \cite{fiore03} and whose X-ray spectra were analyzed by \cite{perola04}.
      Circles and squares identify type 1 and type 2 AGNs according to the
      optical classification (see text for details). Solid and empty symbols
      indicate sources with X/O$>$10 and X/O$<$10, respectively.
      The region matching the definition of obscured, type 2 QSO is delimited
      by the dashed box.
              }
         \label{fig_lxnh}
   \end{figure}

From the (rest-frame) optical point of view, only Abell2690\#029 presents
the classical emission line properties of (dust-) obscured AGN (a ``classical''
type 1.9 spectrum). The other two do not present the classical emission lines
of type 2 AGNs, one having a LINER-like spectrum and the other one being nearly
featureless. However, such spectral properties are sometimes
observed even in local heavily obscured AGNs \citep{maiolino03}.
Probably, in these kind of AGNs the obscuring medium surrounds
completely the nuclear source (``buried'' AGN),
therefore preventing ionizing photons to escape and form
of a Narrow Line Region. Other cases of high-z X-ray sources with
counterparts dominated by the
host galaxy and with only one emission line (H$\alpha$) have been found
by \cite{severgnini03}, but in most of these cases the weakness of AGN
signatures is ascribed to an intrinsic faintness of the AGN
relative to the host galaxy.

Fig.~\ref{fig_lxnh} shows a
$\rm N_H-L_X$ diagram, where we compare the
three sources investigated in this paper (large symbols)
with the HELLAS2XMM
sources identified through optical spectroscopy by \cite{fiore03} and
analyzed, for what concerns the X-ray spectral properties, by
\cite{perola04} (small symbols).  Circles and squares identify
type 1 and type 2 AGN, respectively,
according to the optical spectrum
\citep[here we include in the ``type 2'' class
all those sources without broad emission lines, i.e. including ``Emission
Line Galaxies'' and ``Early Type Galaxies'', see][for a detailed
discussion]{fiore03}. The dashed rectangle indicates the location of
sources conventionally defined as obscured QSO \citep[e.g.][]{mainieri02}.
Although the work by \cite{fiore03} and \cite{perola04}
have delivered a number of new QSO2, only $\sim$10\% of their
whole spectroscopic
sample can be safely identified as QSO2. Conversely, all three ``super-EROs''
identified is the current work
are well within the region of QSO2.

In Fig.\ref{fig_lxnh} we distinguish sources with high and
low X/O ratios (X/O$>$10
and X/O$<$10) by marking them through solid and empty symbols
\citep[a similar plot, $\rm N_H$ vs. X/O, was shown by][]{comastri04}. It
is interesting to note that most sources with X/O$>$10 are characterized
by $\rm L_X > 10^{44}~erg~s^{-1}$ and half of them are absorbed (i.e. QSO2).
On the opposite the low luminosity, Sy-like region is mostly populated
by sources with X/O$<$10 and mostly unabsorbed.
This result indicates that selecting sources with high X/O
is an efficient way to find obscured
QSOs at high redshift, as already pointed out
by \cite{fiore03}. Our findings indicate that by combining this
selection technique with the extremely red colors (EROs) the selection
efficiency of QSO2 may improves even further, as already
suggested by previous studies \citep[][]{brusa05,severgnini05}.
In particular, our results suggest
that by selecting extreme values of X/O ($>$40) and extreme values
of R--K ($>$6) the QSO2 selection efficiency may approach 100\%.

The physical interpretation of our results is relatively straightforward.
On the one hand
the high X/O tends to select obscured AGNs, since obscuration affects
preferentially the optical with respect to the hard X-rays. The redshift
further moves this selection in favor of obscured systems, both because
the observed optical emission samples the UV rest-frame radiation
and because the observed X-ray radiation
samples the harder emission.
On the other hand the
very red colors favor the selection of high-z objects,
especially when the AGN emission is completely absorbed and the optical/near-IR
radiation is therefore dominated by the host galaxy. At least at our
X-ray flux limits  high
redshifts translate also into high X-ray (intrinsic) luminosities.

We note that our results are generally consistent with
previous near-IR spectroscopic studies of optically faint
sources,
both in terms of redshift range and in terms of source characteristics:
most of them are identified as obscured AGNs at high
redshift \citep[][]{gandhi02,willott03,
willott04,severgnini05}.
Altogether, our and previous works indicate that near-IR spectroscopy
is a powerful tool to identify the redshift and the nature of
optically faint sources.

\subsection{Very massive galaxies traced by super-EROs}

Two of the sources spectroscopically identified by us,
BPM16274\#069  and Abell2690\#075, are dominated
by the host galaxy light in the K band (Tab.~1). Thanks to the
spectroscopic redshift we can determine the stellar luminosity 
the host galaxies, as listed in Tab.~3. The inferred luminosities
are $\rm L_K \sim 10^{12}~L_{\odot}$, i.e.
about one order of magnitude higher than local $\rm L_K^*$.
The inferred stellar masses are well in excess of
$\rm 10^{11}~M_{\odot}$ in both cases. A similar result was
obtained by \cite{severgnini05} through the near-IR
spectroscopic identification of an X-ray ERO source with high
X/O and bright in the K band, i.e. with properties very similar
to our sources. These results indicate that the near-IR spectroscopic
identification of EROs X-ray sources with high X/O is an efficient method
also to find very massive galaxies at high redshifts, and therefore to
investigate and observationally test the scenarios of co-evolution
between massive spheroids and black hole accretion
\citep{granato04,dimatteo05}.


If the relation between near-IR
bulge stellar light and Black Hole mass
\citep{marconi03} holds also at high redshift, then our result
would imply that these two sources host black holes with masses
of $\rm 2-3~ 10^9 ~M_{\odot}$. If compared with the X-ray
luminosity, and assuming a bolometric correction factor
$\rm L_{bol}/L_X=30$, such a result would imply that these obscured
QSOs are radiating at a fraction of about 0.04--0.10 of their Eddington
luminosity.
These values are significantly lower than the average $\rm L/L_{Edd}\sim 0.5$
inferred by \cite{marconi04} from the comparison of the high-z X-ray
luminosity functions with the local Black Hole relic mass density. 
Although the uncertainties are large, our results would suggest that
these very massive Black Holes
have already passed their rapidly accreting phase
and are reaching their final masses with lower accretion rates. This
is also a trend which is expected to happen at z$\sim$1--2 for very
massive Black Holes according to the analysis of \cite{marconi04}, as
illustrated in Fig.8 of that paper.
Our results are also consistent with the decresing
$\rm L/L_{Edd}$ at low redshift obtained by \cite{mclure04}.

Finally, the finding of very massive Black Holes, hosted within very
massive (quiescent) galaxies and with low accretion rate
is also consistent with the feedback models for the coevolution of
QSO and spheroids \cite{granato04}. Indeed such models expect that
in the QSO phase most of the star formation and black hole growth
has already occurred, while further accretion is quenced by the feedback
from the QSO itself.

Similar results on $\rm L/L_{Edd}$ were obtained by \cite{comastri04}
and by \cite{brusa05} on a
sample of spectroscopically identified EROs, although less massive
than the sample investigated in this paper.

\begin{table}[bh!]
\label{tab3}      
\caption{Rest-frame properties inferred for the two sources dominated by stellar light.}
\centering          
\begin{tabular}{l c c c c }    
\hline\hline       
Source  &  $\rm L_K$ & $\rm M_{star}$ & $\rm M_{BH}^a$ & $\rm L/L_{Edd}^b$\\ 
        & $\rm (L_{\odot})$ & $\rm (M_{\odot})$ & $\rm (M_{\odot})$ & \\ 
\hline                    
BPM16274\#069  & $\rm 8.3~10^{11}$ & $\rm 4~10^{11}$ & $\rm 2~10^9$ & 0.04 \\ 
Abell2690\#075 &  $\rm  1.7~10^{12}$   & $8~10^{11}$ & $\rm 3~10^9$ & 0.10 \\ 
\hline                  
\end{tabular}
\\Notes: $^a$ Inferred black hole mass assuming the $\rm L_K-M_{BH}$ relation
obtained by \cite{marconi03} for local galaxies. $^b$ Eddington ratio assuming
a bolometric correction $\rm L_{bol}/L_X=30$.
\end{table}


\section{Conclusions}

We have reported the results of a pilot program of near-IR
spectroscopy aimed at identifying four X-ray
sources characterized by extremely high X-ray--to--optical
ratios (X/O$>$40), extremely red colors (6.3$<$R--K$<$7.4)
and relatively bright infrared magnitudes (17.6$<$K$<$18.3).
The optical spectroscopic identification of these sources is
very difficult due to the extremely faint or undetected
optical counterparts. On the contrary,
the near-IR spectroscopic identification
results to be a relatively successful technique, allowing us to
secure the redshift and spectroscopic classification 
of at least two sources, and possibly even a third one,
even with modest integrations times at VLT (40 min per band). The
only source for which spectroscopic constrains cannot be inferred
is the one which was observed in one single band (H).

One object has a red, type 1.9 AGN spectrum (broad H$\alpha$, but not
broad H$\beta$), at a redshift z=2.03. The second one has a
LINER-like emission spectrum, at a redshift z=1.36. The third one
has only a faint emission line, which we tentatively identify with
H$\alpha$ at z=2.13.
At these redshifts the shape of the X-ray spectrum indicates the
presence of gas absorption along the line of sight with column
densities of  $\rm 2\times 10^{22}<N_H<4\times 10^{23}~cm^{-2}$.
The spectroscopic redshifts also imply intrinsic X-ray luminosities in
the range $\rm 4\times 10^{44} < L_X <1.5\times 10^{45}~erg~s^{-1}$, i.e. in the
QSO luminosity range.


Our results corroborate previous studies suggesting that a selection
criterion based on the high X/O and red R--K colors provide an
efficent way to select absorbed QSOs at high redshift.
Our study further suggests that by pushing these criteria to extreme values
(X/O$>$40 and R--K$>$6) the selection efficiency of QSO2s
is probably close to 100\%.


Two of the sources for which we could determine the spectroscopic redshift
are dominated by the host galaxy light in the K band. The inferred
stellar light is about $\rm 10^{12}~L_{\odot}$, about an order of
magnitude higher than local $\rm L_K^*$ galaxies. The inferred
stellar masses are well in excess of $\rm 10^{11}~M_{\odot}$.
These obscured QSOs are therefore hosted in quiescent, very massive galaxies,
already fully assembled even at these high redshifts.

By exploiting the relation between stellar light and black hole mass we
infer that these galaxies host black holes with masses higher than
$\rm 10^9~M_{\odot}$ and that the QSOs radiate at less than
10\% of the Eddington luminosity.
This result suggests that these very massive black holes
have already passed their rapidly accreting phase
and are reaching their final masses with lower accretion rates.
The finding of very massive black holes, hosted within very
massive (quiescent) galaxies and with low accretion rates
is consistent with the feedback models for the co-evolution of
QSO and spheroids.

Summarizing, our results
suggest that the near-IR spectroscopic identification of X-ray sources
with EROs colors and high X/O is a promising method to find absorbed QSOs
and very massive galaxies at high redshift, and therefore to investigate
the coevolution of spheroids and black hole growth.


\begin{acknowledgements}
We thank the anonymous referee for useful comments.
PS acknowledges a research
fellowship from the National Institute for Astrophysics (INAF).
      This work was partially supported by the Italian Ministry
      for Univerisity (MIUR) through grant Cofin-03-02-23 e
      and by INAF
      through grant PRIN/INAF/2003/270.
\end{acknowledgements}


\begin{thebibliography}{}

\bibitem[Barger et al.(2003)]{barger03} Barger, A.~J., et al.\ 
2003, \aj, 126, 632


\bibitem[Brandt et al.(2001)]{brandt01} Brandt, W.~N., et al.\ 
2001, \aj, 122, 2810

\bibitem[Brusa et al.(2005)]{brusa05} Brusa, M., et al.\ 2005, 
\aap, 432, 69 

\bibitem[Comastri \& Fiore (2004)]{comastri04} Comastri, A., Fiore, F. 2004
in Baryons on Cosmic Structures, Ap\&SS 294, 63 

\bibitem[Di Matteo et al.(2005)]{dimatteo05} Di Matteo, T., 
Springel, V., \& Hernquist, L.\ 2005, \nat, 433, 604 

\bibitem[Fiore et al.(2003)]{fiore03} Fiore, F., et al.\ 2003, 
\aap, 409, 79 

\bibitem[Gandhi et al.(2002)]{gandhi02} Gandhi, P., Crawford, 
C.~S., \& Fabian, A.~C.\ 2002, \mnras, 337, 781

\bibitem[Gandhi et al.(2004)]{gandhi04} Gandhi, P., Crawford, 
C.~S., Fabian, A.~C., \& Johnstone, R.~M.\ 2004, \mnras, 348, 529 

\bibitem[Gaskell et al.(2004)]{gaskell04} Gaskell, C.~M., 
Goosmann, R.~W., Antonucci, R.~R.~J., \& Whysong, D.~H.\ 2004, \apj, 616, 
147

\bibitem[Giacconi et al.(2002)]{giacconi02} Giacconi, R., et al.\ 
2002, \apjs, 139, 369

\bibitem[Granato et al.(2004)]{granato04} Granato, G.~L., De 
Zotti, G., Silva, L., Bressan, A., \& Danese, L.\ 2004, \apj, 600, 580


\bibitem[Hasinger et al.(2001)]{hasinger01} Hasinger, G., et al.\ 
2001, \aap, 365, L45

\bibitem[Hopkins et al.(2004)]{hopkins04} Hopkins, P.~F., et al.\ 
2004, \aj, 128, 1112

\bibitem[Koratkar et al.(1995)]{koratkar95} Koratkar, A., Deustua, 
S.~E., Heckman, T., Filippenko, A.~V., Ho, L.~C., \& Rao, M.\ 1995, \apj, 
440, 132

\bibitem[La Franca et al.(2005)]{lafranca05} La Franca, F., et al.
2005, ApJ, submitted

\bibitem[Mainieri et al.(2002)]{mainieri02} Mainieri, V., 
et al. \ 2002, \aap, 393, 425

\bibitem[Mainieri et al.(2005)]{mainieri05} Mainieri, V., et al.\ 
2005, \aap, 437, 805

\bibitem[Maiolino et al.(1996)]{maiolino96} Maiolino, R., Rieke, 
G.~H., \& Rieke, M.~J.\ 1996, \aj, 111, 537

\bibitem[Maiolino et al.(2001a)]{maiolino01a} Maiolino, R., Marconi, 
A., \& Oliva, E.\ 2001a, \aap, 365, 37 

\bibitem[Maiolino et al.(2001b)]{maiolino01b} Maiolino, R., et al.\ 2001b, \aap, 365, 28

\bibitem[Maiolino et al.(2003)]{maiolino03} Maiolino, R., et al.\ 
2003, \mnras, 344, L59

\bibitem[Marconi \& Hunt(2003)]{marconi03} Marconi, A., \& Hunt, 
L.~K.\ 2003, \apjl, 589, L21

\bibitem[Marconi et al.(2004)]{marconi04} Marconi, A., Risaliti, 
G., Gilli, R., Hunt, L.~K., Maiolino, R., \& Salvati, M.\ 2004, \mnras, 
351, 169

\bibitem[McLure \& Dunlop(2004)]{mclure04} McLure, R.~J., \& 
Dunlop, J.~S.\ 2004, \mnras, 352, 1390 

\bibitem[Mignoli et al.(2004)]{mignoli04} Mignoli, M., et al.\ 
2004, \aap, 418, 827


\bibitem[Norman et al.(2002)]{norman02} Norman, C., et al.\ 
2002, \apj, 571, 218

\bibitem[Osterbrock \& Martel(1993)]{osterbrock93} Osterbrock, 
D.~E., \& Martel, A.\ 1993, \apj, 414, 552 

\bibitem[Perola et al.(2004)]{perola04} Perola, G.~C., et al.\ 
2004, \aap, 421, 491

\bibitem[Pickles(1998)]{pickles98} Pickles, A.~J.\ 1998, \pasp, 
110, 863

\bibitem[Rigby et al.(2005)]{rigby05} Rigby, J.~R., Rieke, 
G.~H., P{\' e}rez-Gonz{\' a}lez, P.~G., Donley, J.~L., Alonso-Herrero, A., 
Huang, J.-S., Barmby, P., \& Fazio, G.~G.\ 2005, \apj, 627, 134

\bibitem[Severgnini et al.(2003)]{severgnini03} Severgnini, P., et 
al.\ 2003, \aap, 406, 483 

\bibitem[Severgnini et al.(2005)]{severgnini05} Severgnini, P., et 
al.\ 2005, \aap, 431, 87 

\bibitem[Spergel et al.(2003)]{spergel03} Spergel, D.~N., et al.\ 
2003, \apjs, 148, 175

\bibitem[Szokoly et al.(2004)]{szokoly04} Szokoly, G.~P., et al.\ 
2004, \apjs, 155, 271


\bibitem[Treister et al.(2005)]{treister05} Treister, E., et al.\ 
2005, \apj, 621, 104

\bibitem[Ueda et al.(2003)]{ueda03} Ueda, Y., Akiyama, M., 
Ohta, K., \& Miyaji, T.\ 2003, \apj, 598, 886 

\bibitem[Willott et al.(2003)]{willott03} Willott, C.~J., et al.\ 
2003, \mnras, 339, 397

\bibitem[Willott et al.(2004)]{willott04} Willott, C.~J., et al.\ 
2004, \apj, 610, 140

\end{thebibliography}
\end{document}